\documentclass[amsmat,amssymb,amsfonts,aps,prl,twocolumn,showpacs]{revtex4}

\usepackage{graphicx}
\usepackage{dcolumn}
\usepackage{bm}
\begin{document}

\title{Emergence of half-metallicity in suspended NiO chains}
\author{David Jacob}\email{david.jacob@ua.es}
\author{J. Fern\'andez-Rossier}
\author{J. J. Palacios}
\affiliation{Departamento de F\'isica Aplicada and Instituto Universitario de
Materiales de Alicante (IUMA), Universidad de Alicante, 03690 San Vicente del
Raspeig, SPAIN }

\date{\today}

\begin{abstract}
Contrary to the antiferromagnetic and insulating character of bulk NiO, 
one-dimensional chains of this material can become half-metallic due to the
lower coordination of their atoms. Here we present  \emph{ab initio} electronic
structure and quantum transport calculations of ideal infinitely long  NiO
chains and of more realistic short ones suspended between Ni electrodes.
While infinite chains are insulating, short suspended chains are 
half-metallic minority-spin conductors which display very large magnetoresistance
and a spin-valve behaviour controlled by a single atom.
\end{abstract}

\maketitle


In going from bulk to lower dimensions material properties often change drastically. 
A recent example is that of interfaces between different insulators 
which can actually be metallic\cite{Ohtomo:nature:04}. 
Even more recently, it has been predicted theoretically that certain oxygen surfaces of 
some insulating ceramic oxides can exhibit magnetism and half-metallicity
\cite{Gallego:jpcm:05}. 
The ultimate limit in this respect can be found in atomic chains formed in metallic
nanocontacts which allow to study the transport properties of one-dimensional
systems of atomic size\cite{Agrait:pr:03}. Due to the lower coordination of the atoms
the properties of metallic atomic chains formed in nanocontacts can be remarkably 
different from those in the bulk. For example, Pt nanocontacts can exhibit magnetic 
order when atomic chains are formed\cite{Fernandez-Rossier:prb:05}.
More complex one-dimensional systems like carbon-cobalt atomic chains\cite{Durgun:epl:06} or
organometallic benzene-vanadium wires\cite{Maslyuk:05} have even been predicted to be 
half-metallic conductors.

However, not all metals form atomic chains in nanocontacts, although recently, it has been found 
that the presence of oxygen favours their formation\cite{Thijssen:prl:06}.
For example, experiments with Ni nanocontacts
\cite{Garcia:prl:99,Viret:prb:02}
have never shown evidence of chain formation.
Nevertheless, the presence of oxygen in the contact region could possibly lead to the formation of NiO chains.  
In this context it has also been proposed that the rather moderate magnetoresistive properties 
of pure Ni nanocontacts \cite{Jacob:prb:05,Bagrets:05} could be enhanced considerably 
by the presence of oxygen adsorbates on the surface of the Ni electrodes\cite{Papanikolaou:jpcm:03}.
On the other hand, bulk NiO is a common example of a correlated insulator with antiferromagnetic (AF) 
order (see, e.g., Ref. \onlinecite{Sawatzky:prl:84,Moreira:prb:02}), which remains insulating
even above the N\'eel temperature when the AF order is lost. Thus it is not at all obvious 
whether or not oxidized Ni nanocontacts or NiO chains should be conductors.  

In this Letter we investigate the electronic and magnetic structure and the
transport properties of one-dimensional NiO chains, both idealized infinite
ones and more realistic short ones suspended between Ni nanocontacts.
Anticipating our most important results our \emph{ab initio} quantum transport
calculations show that short NiO chains suspended between Ni nanocontacts can
become half-metallic conductors, i.e.,  carry an almost 100\%
spin-polarized current. This result holds true even for a single O atom in
between Ni electrodes. Consequently, for antiparallel  alignment of the
electrode magnetizations the transport through  the contact is strongly
suppressed resulting in very large MR [difference in resistance 
between antiparallel and parallel alignment of the magnetizations 
of the electrodes normalized either to the higher resistance value (${\rm MR}_1$)
or to the lower one (${\rm MR}_2$)]: ${\rm MR}_1 \approx 90\%$ 
and  ${\rm MR}_2 \approx 700\%$, respectively.

The electronic structure of both  idealized infinitely long
NiO chains and the more realistic short suspended chains have been calculated in
the density functional theory (DFT) approximation using the hybrid functional
B3LYP\cite{Becke:jcp:93}, which combines local and non-local exchange. This 
functional has been successful in describing the electronic and magnetic structure
of some strongly correlated materials like bulk NiO. In particular, it describes 
very well the charge-transfer character, the magnitude of the gap and the magnetic 
moment of NiO, and predicts the correct AF order\cite{Moreira:prb:02}.
It is important to stress here that due to the insufficient cancellation of
the self-interaction in the local exchange functional of the local density
approximation (LDA) the occupied narrow  $3d$ bands are raised in energy. As a
result LDA severerly underestimates the gap of bulk NiO \cite{Leung:prb:91}.
The generalized gradient approximation (GGA) to the exchange 
functional improves somewhat the description of NiO but the 
energy gap of $\sim$2eV is still too small and also the charge-transfer 
character is not captured correctly \cite{Leung:prb:91}.
B3LYP corrects the self-interaction error inherent in LDA and GGA
by mixing the exact non-local Hartree-Fock (HF) exchange with the
GGA exchange functional.
The  DFT-based alternatives that correct the self-interaction error like LDA+U, 
the self-interaction-corrected LDA (SIC-LDA), and the GW approximation 
lead to similiar results as those obtained from B3LYP\cite{Moreira:prb:02}.

The electronic structure of infinite NiO chains has been calculated with the
CRYSTAL03 program package  \cite{Crystal:03} while for the electronic structure
and quantum transport calculations of the suspended  chains we have used our
\emph{ab initio} quantum transport package ALACANT\cite{palacios:prb:02,ALACANT}. The infinite 
chain calculations were done with elaborate all-electron basis sets for Ni and O 
\cite{Doll:ss:03,Ruiz:jssc:03}, similiar to those employed for reported HF and B3LYP 
calculations of bulk NiO\cite{Moreira:prb:02}, but extended with a diffusive 
$sp$-function in the case of Ni and a $d$-polarization function in the case of O. 
Using these basis sets we reproduce the B3LYP results for bulk NiO \cite{Moreira:prb:02}. 
We have also employed minimal basis sets with effective core pseudo-potentials 
described in earlier work \cite{Jacob:prb:05} for both Ni and O, and have found 
that the results change very little. Thus we have employed these minimal basis 
sets for the more demanding transport calculations.

The electronic properties of NiO are, to a large extend, determined by the
atomic scale properties, like the crystal field splitting of the Ni and O
energy levels and the amount of electron charge transfered from Ni to O.
The latter, in turn, is determined by the interplay between Madelung binding 
energy, the ionization potential of Ni, and the electron affinity of O. 
Due to the lower coordination and the corresponding decrease in Madelung 
binding energy the electron transfer from Ni to O is less favourable in an
atomic chain than in bulk (where the electron transfer is almost complete 
resulting in an ionic configuration of Ni$^{2+}$O$^{2-}$).
The proper starting point to discuss the formation of energy  bands in the
one-dimensional NiO chain are therefore the univalent ions Ni$^+$ and O$^-$.

In order to understand how the low coordination affects the atomic properties of the 
constituting ions we have first performed B3LYP calculations of both a single Ni$^+$ ion 
and a single O$^-$ ion each in the field of point charges that mimic the crystal field 
of a one-dimensional chain of univalent Ni and O ions.
\begin{figure}
  \includegraphics[width=0.6\linewidth]{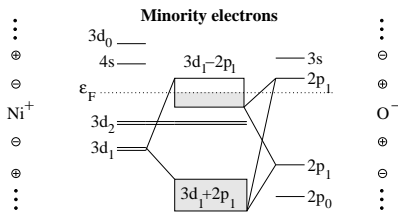}
  \caption{Schematic one-electron energies of a one-dimensional NiO chain in z-direction
    for minority-spin. To the left and right the orbital energies of an 
    individual Ni$^+$ cation and O$^-$ anion in the crystal field of a one-dimensional 
    Ni$^+$O$^-$ chain are shown. In the center the formation of valence and conduction 
    bands by hybridization of Ni $3d$ and O $2p$ orbitals is shown.
  }
  \label{fig:scheme}
\end{figure}
We find that for both ions the spin-doublet state ($S=1/2$) minimizes the energy
as is the case for the free ions. In Fig. \ref{fig:scheme} we show schematically 
the energy levels of Ni$^+$ and O$^-$ in the presence of the point charges
for minority spin only.
Interestingly, for minority spins, 
the occupied Ni $3d_1$ orbitals ($d_{xz}$ and $d_{yz}$) of the Ni$^+$ 
ion fall energetically in between the occupied and  unoccupied O $2p_1$ orbital 
($p_x$ and $p_y$) of the O$^-$ ion. 
Thus the $3d_1$ and $2p_1$ orbitals can form two filled degenerate bonding bands and 
two degenerate partially filled antibonding bands as indicated in the middle part of
Fig. \ref{fig:scheme}.  The $3d_2$ ($d_{xy}$ and $d_{x^2-y^2}$) doublet is somewhat 
above in energy to the $3d_1$ but cannot hybridize with the oxygen $2p$ orbitals. 
The Ni $3d_0$ ($d_{3z^2-r^2}$) and $4s$ are empty while the O $2p_0$ ($p_z$) and 
$2s$ orbitals are  filled and much lower in energy so that no hybridization takes 
place though symmetry would allow for it.

On the other hand, for majority-spin electrons (not shown) all five Ni $3d$
orbitals are filled while the $4s$ is also  empty, i.e., the Ni$^+$ valence
configuration is $3d^9$ and not $4s^1 3d^8$ as for the free Ni$^+$ ion. 
Moreover, all of the O $2p$ orbitals are filled so that the Ni$^+$ and O$^-$
ions can only form  either completely filled or completely empty bands for the
majority spin.  Thus the ionic picture suggests that a
one-dimensional NiO chain should become a half-metallic conductor where only one 
kind of spin levels forms conducting bands. In this case only the minority-spin
$3d_1$ bands would conduct so that, according to the proposed 
classification in Ref. \onlinecite{Coey:jphysd:02}, it would be of Type I$_{\rm B}$.

\begin{figure}
    \includegraphics[width=\linewidth]{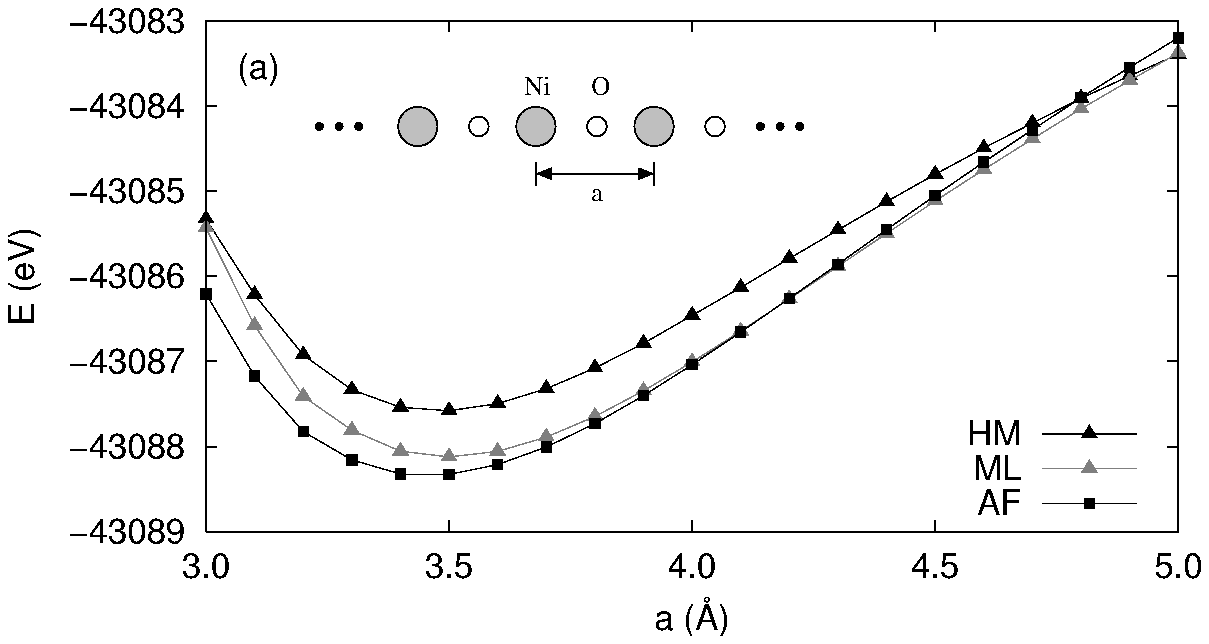}
  \begin{tabular}{cc}
    \includegraphics[width=0.49\linewidth]{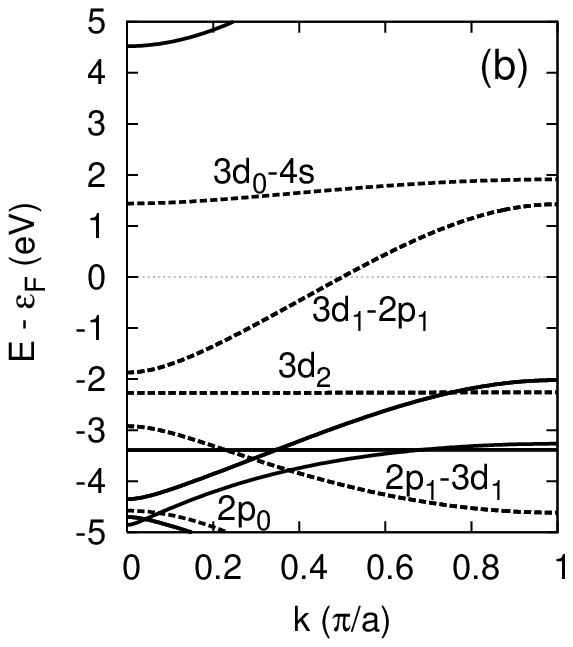} &
    \includegraphics[width=0.49\linewidth]{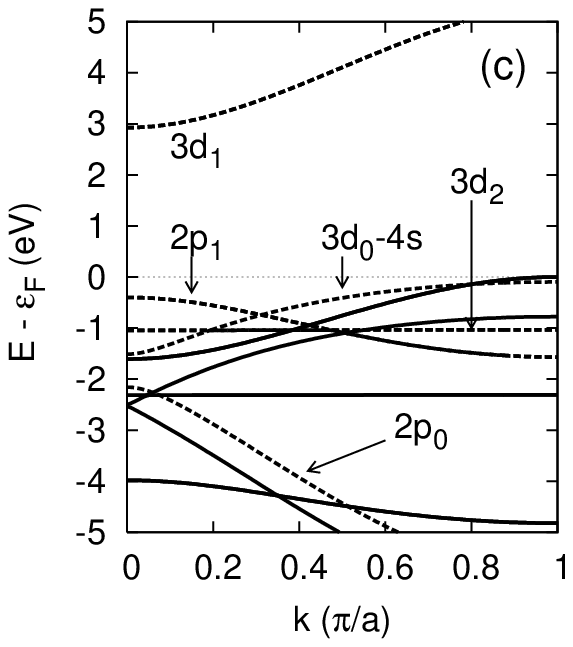} 
  \end{tabular}
  \caption{ (a) Energy per unit cell of infinite NiO chain in dependence 
    of lattice spacing $a$. Black triangles indicate the half-metallic state 
    (HM), grey triangles the molecule-like insulating state 
    with FM order (ML), and black boxes the insulating state with AF order (AF).
    (b) Band structure of HM state for a lattice 
    spacing of 3.6\r{A}. Solid lines indicate majority-spin bands and 
    dashed lines indicate minority-spin bands. 
    (c) Same as (b) but for ML FM state.
  }
  \label{fig:chain}
\end{figure}

Not surprisingly, our calculations for inifinite one-dimensional NiO chains
(Fig. \ref{fig:chain}) show that a  univalent ionic configuration as an initial
guess results in a half-metallic state for large separation
of the individual chain atoms (i.e., large lattice spacing of 5\r{A}) as
suggested by the ionic picture. However, as can be seen from
Fig.\ref{fig:chain}(a), the half-metallic state is only a metastable state for
most values of the lattice spacing.
This half-metallic state  is ``shadowed'' by a second state with FM  order and
insulating character. By successively decreasing the lattice spacing $a$ of the
chain and using the  (half-metallic) state of the previous step for the initial
guess, the half-metallic state can be generated also for smaller  inter-atomic
distances which points towards its metastability. 
Around the equilibrium lattice spacing ($a \sim 3.4$\r{A}) the ground state of the chain has AF order and is 
of insulating character with a substantial gap of $\sim 4$eV like in bulk. When stretched out of equilibrium 
the FM state and the AF state become comparable in energy until finally, at a lattice spacing of $\sim 4.2$\r{A}, 
the FM state becomes the ground state. 

The band structure diagram in Fig. \ref{fig:chain}(c) shows that the metastable
state with FM order corresponds indeed to the half-metallic state suggested by
the ionic picture: The half-filled doubly-degenerate conduction band is formed by minority-spin
Ni $3d_1$ orbitals hybridized with O $2p_1$ orbitals while the Ni $3d_2$
orbitals do  not hybridize with O $2p$ orbitals and thus form a flat valence
band. The lowest-lying empty band is formed by the  minority-spin Ni $3d_0$
orbital which is slightly hybridized with the Ni $4s$ orbital. On the other
hand the stable state with FM order and insulating character (see band
structure in Fig. \ref{fig:chain}(b))  actually corresponds to the ground state
of the NiO molecule which is a ${}^3\Sigma^{-1}$ state\cite{Doll:prb:97}. The 
main difference with the half-metallic state is that now the
\emph{non-degenerate} Ni $3d_0$ and $4s$ orbitals  form a minority-spin valence
band while the minority-spin \emph{doubly degenerate half-filled} antibonding
band  composed of Ni $3d_1$ and O $2p_1$ bands is emptied and a substantial gap
of $\sim 3$eV opens. Thus the infinite chain  behaves like an insulator for
reasonable values of the chain stretching. 

\begin{figure}
  \includegraphics[width=\linewidth]{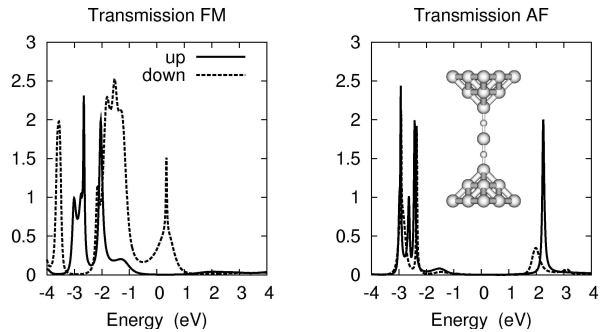}  
  \caption{
    Transmission per spin channel for suspended chain consisting of a O-Ni-O bridge and the
    Ni tip atoms shown in inset of right panel in the case of FM order (left) and of AF order 
    (right). For AF order the magnetization of the Ni atom in the center of the chain is 
    reversed with respect to the two Ni tip atoms. Distance between a Ni tip atom and the 
    center atom is 3.6\r{A}.
  }
  \label{fig:transm2}
\end{figure}

Atomic chains formed in break junctions have a finite length and are suspended 
between electrodes. It is well known that the
contact between the atomic chain and the electrode tip will have considerable
effect on  the electronic structure of the chain, especially when $d$-orbitals
are involved like is the case here\cite{Jacob:prb:05}. 
\begin{figure}
  \includegraphics[width=\linewidth]{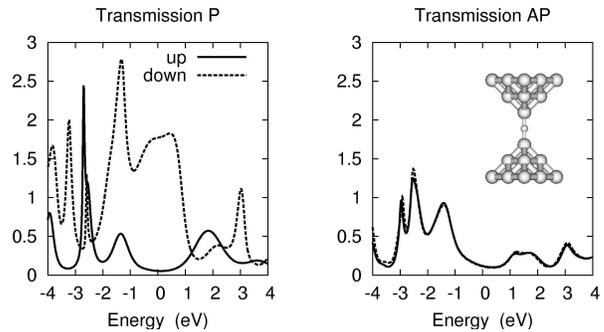}
  \caption{
    Transmission per spin channel in the case of P (left)
    and AP (right) alignment of the electrode magnetizations
    for NiO chain consisting of one oxygen atom bridging the
    two Ni tip atoms of the Ni electrodes as shown on the inset 
    in the right panel. The separation of the two Ni tip atoms is 3.6\r{A}.
  }
  \label{fig:transm}
\end{figure}
We have thus calculated the electronic structure and transport properties
of  both a single oxygen atom and a  O-Ni-O chain bridging the two tips 
of a Ni nanocontact as shown in the insets of the right panels of Fig. 
\ref{fig:transm} and Fig.\ref{fig:transm2}.
In the case of the single oxygen atom (Fig. \ref{fig:transm}) the electron transport 
is almost 100$\%$ spin-polarized around the Fermi level for parallel (P) alignment 
of the magnetizations of the two Ni electrodes. Moreover, an orbital eigenchannel
analysis \cite{Jacob:prb:06} reveals that the  transport is mainly due to two
almost perfectly transmitting minority-spin channels  composed of Ni $3d_1$ and
O $2p_1$ orbitals, i.e. they correspond to the doubly-degenerate conduction
band of the metastable half-metallic state in the  perfect chain. Thus the
half-metallic state which was suppressed in the idealized case  of the infinite
chain emerges in the more realistic situation of a short suspended chain. 
We can understand this phenomenon in terms of the orbital blocking mechanism proposed earlier 
in the context of Ni nanocontacts\cite{Jacob:prb:05,Jacob:prb:06}. The highest minority-spin 
valence band of the insulating state with FM order in the infinitely long chain has a strong 
contribtution from the Ni $3d_0$ orbital which is not ``compatible'' with the geometry of the 
pyramid shaped Ni contacts, so that this band is blocked and thus cannot be occupied. 
Instead, the doubly-degenerate band composed of Ni $3d_1$ orbitals hybridized with
O $2p_1$ orbitals is partially filled resulting in the half-metallic state which in the 
perfect chain is only metastable. Thus the orbital blocking by the contacts actually turns the 
chain into a half-metallic conductor.
Consequently, the conductance is strongly suppressed in the case of antiparallel (AP) alignment 
of the magnetizations of the Ni electrodes as can be seen from the right panel of Fig. 
\ref{fig:transm} and the MR becomes very large: ${\rm MR}_1 \approx 90\%$ and ${\rm MR}_2 \approx 700\%$.

Varying the distance $d$ between the Ni tip atoms leads to similiar results as those shown in Fig. 
\ref{fig:transm}. For the P case the current through the chain is almost 100\% spin-polarized with
two open minority channels composed of Ni $3d_1$ and O $2p_1$ orbitals, while
for the AP case it is strongly suppressed, resulting in very high MR values between 80\% and 90\%
for MR$_1$ for $d$  between 3.0\r{A} and 5.0\r{A}. 
Geometry relaxations for different values of the tip-tip distance show that for small distances
the oxygen atom goes into a zigzag position. The bonding angle decreases with increasing distance
 until it becomes zero at 3.6\r{A}. Finally, the chain breaks for $d > 4.8$\r{A}.
Thus the scattering is strong for small distances $d < 3.6$\r{A} when the bonding angle
is appreciable and for large stretching, $d > 4.2$\r{A}, resulting in a considerable reduction in the
conduction of the two minority channels. 

In longer suspended chains the insulating state with FM order starts to emerge inside the chain and
a away from the contacts. As a result the conductance is reduced as can be seen already in the case of the O-Ni-O 
bridge (Fig. \ref{fig:transm2}). The minority-spin conduction is reduced considerably ($\sim 30\%$) 
compared to the case of the single oxygen bridge. On the other hand the conduction of the majority-spin 
channel becomes practically zero ($<0.2\%$). This can be understood by the fact that the residual 
majority-spin channel conductance of the single oxygen bridge is due to direct hopping of Ni $s$ electrons 
between the electrodes and therefore vanishes when the distance between the electrodes is large. 
The finite minority-spin conductance opens up the possibility to an interesting phenomenon:
When the middle Ni atom reverses its spin, the conductance drops to nearly zero (see right panel of Fig. 
\ref{fig:transm2}) since the AF chain is insulating. In other words, this system behaves as a 
\emph{single atom spin valve} which presents an extremely large MR even higher than that reported above for
the single oxygen bridge:  ${\rm MR}_1 \approx 99\%$ and ${\rm MR}_2 \approx 10,000\%$.
Apart from controlling the magnetization direction of the central atom by a magnetic field, Fig. \ref{fig:chain}(a) 
suggests that a mechanical control of the spin valve (by  stretching the chain) would also be possible.

In conclusion, we have shown that one-dimensional infinite NiO chains are insulating 
and have AF order around the equilibrium lattice spacing but can actually become a FM 
insulator when stretched out of equilibrium, in contrast to bulk NiO. 
In the more realistic case of a short NiO chain suspended
between Ni nanocontacts the chain becomes an almost 100\% spin-polarized conductor. 
The emergence of almost perfect half-metallicity (i.e. almost 100\% spin-poalrization 
of the conduction channel) in suspended chains leads to a strong suppression of the
current for AP alignment of the electrodes resulting in very large MR values
of ${\rm MR}_1 \approx 90\%$ and ${\rm MR}_2 \approx 700\%$, respectively.
This could perhaps explain to some extend the very large MR values in Ni nanocontacts 
obtained in some experiments\cite{Garcia:prl:99} where oxygen is likely to be present. 
Finally, the O-Ni-O bridge suspended between Ni electrodes operates as a 
\emph{single atom spin valve} where the currentflow is
controlled by the magnetization of a single atom.

We thank C. Untiedt, A. J. Per\'ez-Jimenez and J. M. van Ruitenbeek 
for fruitful discussions and J. L. McDonald from GuiriSystems for building 
the Beowulf cluster facility in Alicante and providing us with technical support.
D.J.  acknowledges financial support from MECD under grant UAC-2004-0052.
J.F.R acknowledges financial support from Grant No.FIS200402356 (MCyT), 
No. GV05-152, and Ramon y Cajal Program (MCyT).   
J.J.P. acknowledges financial support from MAT2005-07369.


\bibliography{matcon}

\end{document}